\documentclass[aps,nofootinbib,nobibnotes,noeprint,preprintnumbers,floatfix,twocolumn]{revtex4-2}
\usepackage{graphicx} 
\usepackage{hyperref}
\usepackage[utf8]{inputenc}
\usepackage{graphicx}
\usepackage{amsmath}
\usepackage{xcolor}
\usepackage[caption = false]{subfig}
\usepackage{graphicx}
\usepackage{dcolumn}
\usepackage{bm}
\usepackage{hyperref}
\usepackage{xcolor}
\usepackage{amsmath,amsfonts,amssymb}
\usepackage{commath}
\usepackage{mathtools}
\usepackage{flushend}
\usepackage[utf8]{inputenc} 
\usepackage[T1]{fontenc}

\newcommand{\bs}[1]{\boldsymbol{#1}}
\renewcommand{\d}{\mathrm{d}}

\begin{document}
\preprint{FERMILAB-PUB-24-0365-T}
\title{A New Probe of $\mu$Hz Gravitational Waves with FRB Timing}
\author{Zhiyao Lu$^1$}
\email{luzhiyao@stu.pku.edu.cn}
\author{Lian-Tao Wang$^2$}
\email{liantaow@uchicago.edu}
\author{Huangyu Xiao$^{3,4}$}
\email{huangyu@fnal.gov}
\affiliation{$^1$School of Physics and State Key Laboratory of Nuclear Physics and Technology, Peking University, Beijing 100871, China}
\affiliation{$^2$ Department of Physics, Kadanoff Center for Theoretical Physics \& Enrico Fermi Institute, University of Chicago }
\affiliation{$^3$Kavli Institute for Cosmological Physics, University of Chicago, Chicago, IL 60637}
\affiliation{$^4$ Astrophysics Theory Department, Theory Division, Fermilab, Batavia, IL 60510, USA}

\begin{abstract}
    We propose Fast Radio Burst (FRB) timing, which uses the precision measurements of the arrival time differences of repeated FRB signals along multiple sightlines, as a new probe of gravitational waves (GWs) around nHz to $\mu$Hz frequencies, with the highest frequency limited by FRB repeating period. The anticipated experiment requires a sightline separation of tens of AU, achieved by sending radio telescopes to space.
    We find the signal of arrival time difference induced by GWs depends only on the local GWs in the solar system and we can correlate the measurements from different FRB sources or the same source with different repeaters, which leads to a better sensitivity with a larger number of FRB repeaters detected. The projected sensitivity shows this method is a competitive probe in the nHz to $\mu$Hz frequency range. It can fill the `$\mu$Hz gap' between pulsar timing arrays and Laser Interferometer Space Antenna (LISA) and is complementary to other proposals of GW detection in this frequency band.
\end{abstract}
\maketitle

\section{Introduction}
Since the direct discovery of gravitational waves (GWs) in the Hz to kHz frequencies by LIGO/Virgo \cite{LIGOScientific:2016aoc,LIGOScientific:2017vwq,LIGOScientific:2018mvr,LIGOScientific:2020ibl,KAGRA:2021vkt}, a new era of astronomical observations has commenced, with a new gravitational window that provides insights into the dynamics of massive objects like black holes and neutron stars during their mergers. The increasing evidence of a stochastic GW signal in the nHz band from pulsar timing arrays (PTAs) further advances the frontiers of our exploration \cite{Kramer:2013kea,EPTA:2015gke,Shannon:2015ect,10.1093/mnras/stw347,Aggarwal:2018mgp,Kerr_2020,NANOGrav:2020bcs,NANOGrav:2023ctt,NANOGrav:2023hde,NANOGrav:2023gor,NANOGrav:2023hvm,NANOGrav:2023hfp}, opening up potential opportunities for supermassive black hole mergers \cite{Rajagopal:1994zj,Wyithe:2002ep,Jaffe:2002rt,Sesana:2004sp,Burke-Spolaor:2018bvk,Sato-Polito:2023gym,Sato-Polito:2024lew} and discovering new physics beyond the standard model (BSM) \cite{NANOGrav:2023hvm}, including cosmic inflation \cite{Guzzetti:2016mkm}, scalar-induced GWs \cite{Domenech:2021ztg,Yuan:2021qgz,Ebadi:2023xhq}, first-order phase transitions \cite{Caprini:2007xq,Caprini:2015zlo,Caprini:2019egz,Hindmarsh:2013xza,Cutting:2020nla}, cosmic strings \cite{Vachaspati:1984gt,Hindmarsh:1994re,Accetta:1988bg,Leblond:2009fq,Blanco-Pillado:2017oxo,Blasi:2020mfx,Chang:2021afa,Schmitz:2024hxw}, and domain walls \cite{Hiramatsu:2010yz,Kawasaki:2011vv,Hiramatsu:2013qaa,Saikawa:2017hiv}. Existing and proposed gravitational-wave observations will extensively explore frequency bands from nHz to GHz \cite{Graham:2012sy,Baker:2019nia,lisa2018lisa,amaro2017laser,Armano:2018kix,TianQin:2015yph,Milyukov:2020kyg,Dimopoulos:2007cj,Dimopoulos:2008sv,Hogan:2011tsw,Graham:2017pmn,Coleman:2018ozp,canuel2018exploring,abe2021matter,Tino:2019tkb,Badurina:2019hst,zhan2020zaiga,AEDGE:2019nxb,Badurina:2021rgt,Kolkowitz:2016wyg,phinney2004big,Crowder:2005nr,Seto:2001qf,kawamura2019space,Kawamura:2020pcg,Maggiore:2019uih,collaboration2015instrument,evans2021horizon,Ackley:2020atn,Aggarwal:2020olq,Berlin:2021txa,Baum:2023rwc,Berlin:2023grv} and even higher frequencies\cite{mcdonald2024}. However, the frequency band near $\mu$Hz, between the reaches of the PTAs and Laser Interferometer Space Antenna (LISA), is challenging to probe. 
This is known as the `$\mu$Hz gap'.

There are a few proposals in this frequency band, such as $\mu$Ares \cite{Sesana:2019vho}, astrometric techniques \cite{Wang:2020pmf}, Doppler tracking with the
Uranus orbiter \cite{Zwick:2024hag}, and asteroids \cite{Fedderke:2021kuy}. The $\mu$Ares proposal is similar to LISA, but requires a much longer baseline and improved low-frequency
test-mass isolation \cite{Armano:2018kix}. The most sensitive proposal is likely installing atomic clocks and transmitter/receiver link systems in asteroids, which reduces the acceleration noise due to the heavy mass of asteroids. 
In this work, we propose Fast Radio Burst (FRB) \textit{timing} that uses solar-system scale interferometry as an alternative probe to gravitational waves in the $\mu$Hz gap. Fast Radio Bursts are radio transients with a typical duration of a few milliseconds and are mostly extragalactic origin. While the exact origin of FRBs remains unclear, thousands have been discovered \cite{CHIMEFRB:2021srp,CHIMEFRB:2023hfj}, with more than fifty known to repeat, often on timescales of tens to thousands of hours \cite{CHIMEFRB:2023myn}, which offer a valuable tool for studying various aspects of cosmology \cite{McQuinn:2013tmc,Ravi:2018ose,Prochaska_2019,Heimersheim:2021meu,Li:2017mek,Zitrin:2018let,Wucknitz:2020spz,Tsai:2023tyw,Pearson:2020wxb,Dai:2017twh,Saga:2024dva,Hagstotz:2021jzu} and BSM physics \cite{Munoz:2016tmg,Buckley:2020fmh,Xiao:2022hkl,Prabhu:2023cgb,Lemos:2024jbl,Gao:2023xbi,Wang:2024sdz}.

Unlike pulsars, FRBs are not clocks in themselves, as they do not exhibit a self-similar repeating pattern in their burst profiles to measure local time changes. In other words, FRBs cannot serve as their own reference clocks. However, we can create a reference clock by placing two 10-meter radio dishes in space separated by tens of Astronomical Unit (AU). Because FRB events originate from extremely compact sources, the signals observing from different sightlines remain coherent, which allows us to correlate the FRB signals received from two dishes, providing high-precision measurement on the \textit{arrival time difference} of FRBs at two different detectors. It is possible to achieve a relative timing of the signals with sub-nanosecond precision through coherent analysis of electric-field time series of FRBs \cite{Cassanelli:2021oaw}. A fraction of FRB sources repeatedly emit FRBs, which allows us to measure the arrival time difference induced by local gravitational waves and monitor their temporal changes. 
If the frequency of GWs is much higher than the rate of FRB detection, the measurements between different FRB events are not correlated and we lose the sensitivity.
Therefore, the repeating or detection rate of FRBs, which is often on timescales of tens to thousands of hours \cite{CHIMEFRB:2023myn}, determines the upper limit on the frequency band of gravitational waves that can be probed with FRB timing, which produces an interesting frequency band that falls into the $\mu$Hz gap.

\begin{figure}
    \centering
    \includegraphics[width=0.6\linewidth]{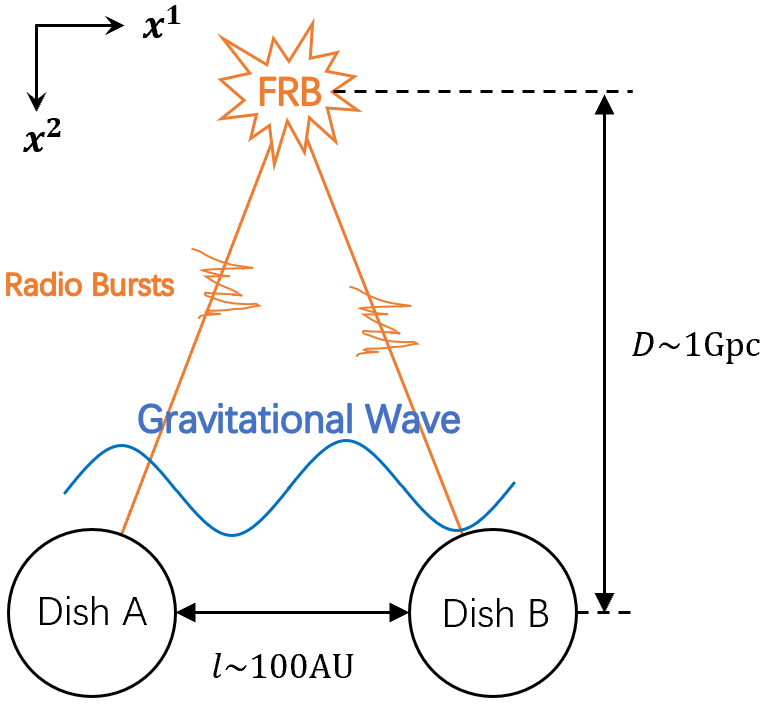}
    \caption{A cartoon picture of FRB timing with two radio dishes in the solar system separated by $l\sim 100$ AU. Both dishes will observe the same FRB source from different sightlines. One sightline can be the reference clock of the other by correlating the burst profile and the arrival time difference of FRBs at two dishes will be the measurement.
  The gravitational wave propagating in the $x$ direction will induce temporal changes on the arrival time difference, which can be probed with FRB repeaters.}
    \label{fig:setup}
\end{figure}

While the space mission to achieve FRB timing is ambitious and challenging, solar-system scale interferometry on FRBs was proposed to measure cosmological distances with a sub-percent precision \cite{Boone:2022pdz} and dark matter substructures such as axion miniclusters \cite{Xiao:2024qay}. Our work on the gravitational waves will further strengthen the scientific case for this potential mission. Additionally, FRB timing can often avoid astrophysical systematics that are relevant to pulsar timing. For example, timing-varying electron density (or dispersion measure) induced by the solar wind will greatly affect the timing precision of pulsars, which can be fitted away with a broad frequency band of observing FRBs. This is because FRB timing uses the correlation of the electric field at different detectors and the dispersion measure is only a phase in the correlation. The detectors need to be localized with an accuracy of several centimeters, which can be achieved in the outer solar system with weekly calibration, and the satellites also need an atomic clock with accuracy comparable to the Deep Space Atomic clock \cite{Boone:2022pdz}.
We will show with our calculation that the FRB timing directly measures the local gravitational waves in the solar system and is insensitive to the GW at FRB source. Therefore, every time an FRB is observed at both detectors, we have one measurement of the arrival time difference which manifestly depends on local gravitational waves, regardless of the FRB source from which it comes.
This allows us to correlate signals from different FRB sources and reach a better sensitivity with more FRBs detected. The total number of FRB events detected per unit time is the only relevant parameter for the sensitivity of FRB timing. There are different search strategies for our space telescopes to maximize the detection rate. One is to point the telescope to a known FRB source that is bright and has a high repeating rate. Another might involve monitoring the active phase of different repeating sources on Earth and allocating the observation time in space accordingly. We will show a conservative burst rate of 100 per year already gives a competitive projected sensitivity while future explorations on better strategies might further improve it.

\section{Gravitational Waves Signals in FRB Timing}
 
In the following, we will show the arrival time difference in two dishes induced by gravitational waves only depends on local GWs in the solar system.
Our main result in this section is presented in Eq.~\ref{equ:delta_t}, which gives the arrival time difference as a function of the baseline of the space interferometer, local GW strain, and frequency. For similar calculations on the time delays induced by GWs in the strong lensing case, see Ref.~\cite{Allen:1989my,Allen:1990ta,Frieman:1994pe}. 
For a simple estimation, we first consider a monochromatic plane GW with a frequency of $\omega_g$ propagating in the $x$ direction. As shown in Fig.~\ref{fig:setup}, the FRB source is at a distance of $D$ ($\sim $Gpc), while the two radio dishes in space are separated by $l$, which is of order 10 AU. The metric perturbation is given by \footnote{We have assumed the GWs are propagating in a flat space. The expansion of the Universe might also induce some time delays comparable to the GW signal we discuss here. However, the dominant contribution from the expanding Universe is either static or linearly depending on measurement time \cite{Li:2017mek,Boone:2022pdz}, which is distinguishable from these oscillatory GW signals.}\cite{Maggiore:2007ulw,Maggiore:2018sht,Creighton:2011zz} 
\begin{equation}\label{equ:metric_pert}
    h_{\mu\nu}=\begin{bmatrix}
        0&0&0&0\\
        0&0&0&0\\
        0&0&h_+&h_\times\\
        0&0&h_\times&-h_+
    \end{bmatrix}\cos(\omega_g x^0-\omega_g x^1+\varphi_0)
\end{equation}
Here $\omega_g$ is the angular frequency of gravitational wave, and $\varphi_0$ is an initial phase. We are interested in the trajectory of photons in the GW background, which is governed by $\frac{dx^\mu}{ds}=V^\mu$. $s$ parametrizes the trajectory, which is taken to be 0 at the source and 1 at the detector.
In the presence of gravitational waves, the four-velocity satisfies the geodesic equation 
\begin{equation}
    \frac{\d V^\mu}{\d s}+\Gamma^\mu_{\nu\rho}V^\nu V^\rho=0
\end{equation}
or equivalently,
\begin{equation} \label{eq:geodesic}
    \frac{\d V^\mu}{\d s}+\eta^{\mu\sigma}\left(\partial_{\rho}h_{\nu\sigma}-\frac{1}{2}\partial_{\sigma}h_{\nu\rho}\right)V^\nu V^\rho=0,
\end{equation}
where $\eta_{\mu\nu}$ is the metric of the Minkowski spacetime. The above expression, with proper boundary conditions, will give the full set of equations that are needed for determining the propagation of photons in GW background and thus the FRB arrival time difference among different detectors. 
Since the strain of gravitational waves satisfies $h\ll1$, we can keep the lowest order only and obtain linearized solutions of the geodesic equation: 
\begin{subequations}
\begin{align}
    \label{equ:wave_vec_1}
    &\begin{aligned}
        V^0(s)=&\sqrt{D^2+\left(\frac{l}{2}\right)^2}+\Delta V^0\\
        &-\frac{h_+}{2}\frac{D^2}{\tilde{D}}\cos(\omega_g \tilde{D} s+\omega_g t_0+\varphi_0)
    \end{aligned}\\
    \label{equ:wave_vec_2}
    &V^1(s)=\frac{l}{2}+\Delta V^1-\frac{h_+}{2}\frac{D^2}{\tilde{D}}\cos(\omega_g \tilde{D} s+\omega_g t_0+\varphi_0)\\
    \label{equ:wave_vec_3}
    &V^2(s)=D+\Delta V^2-Dh_+\cos(\omega_g \tilde{D} s+\omega_g t_0+\varphi_0)\\
    \label{equ:wave_vec_4}
    &V^3(s)=\Delta V^3-Dh_\times\cos(\omega_g \tilde{D} s+\omega_g t_0+\varphi_0)
\end{align}
\end{subequations}
Where $\omega_g$ is the angular frequency of gravitational wave, $s$ is the parametrization of the photon trajectory, and $t_0$ is the time of FRB emission at the source, which only enters the calculation as a phase.
The above solutions of four-velocity apply to photons observed at Dish B as shown in Fig.~\ref{fig:setup}, while the result of Dish A can be easily obtained by reversing the sign of $l$.
$\Delta V^\mu$ are constants of order $\mathcal{O}(h)$ fixed by boundary conditions, and we have defined $\tilde{D}\equiv \sqrt{D^2+\left(\frac{l}{2}\right)^2}-\frac{l}{2}$. Since $D\gg l$, we expect $\tilde{D}$ is very close to the value of $D$. We keep the exact form in the previous expression to figure out the exact expansion of these small terms, as the leading term will cancel when we compare the arrival time difference between dish B and dish A.


Integrating \eqref{equ:wave_vec_1} - \eqref{equ:wave_vec_4} over $[0,1]$, we should reproduce the coordinate difference\footnote{We have chosen the transverse-traceless gauge (TT gauge), where the coordinates of free-falling objects are not changed by the gravitational waves.} between the source and the detector, which imposes the following boundary conditions
\begin{subequations}
\begin{align}
    \label{equ:space_1}&\Delta V^1=\frac{h_+}{2\omega_g}\frac{D^2}{\tilde{D}^2}[\sin(\omega_g \tilde{D}+\omega_g t_0+\varphi_0)-\sin(\omega_g t_0+\varphi_0)]\\
    \label{equ:space_2}&
    \Delta V^2=\frac{h_+}{\omega_g}\frac{D}{\tilde{D}}[\sin(\omega_g \tilde{D}+\omega_g t_0+\varphi_0)-\sin(\omega_g t_0+\varphi_0)]\\
    &\Delta V^3=\frac{h_\times}{\omega_g}\frac{D}{\tilde{D}}[\sin(\omega_g \tilde{D}+\omega_g t_0+\varphi_0)-\sin(\omega_g t_0+\varphi_0)]
\end{align}
\end{subequations}
The null condition of four-velocity $V^\mu V^\nu g_{\mu\nu}=0$ can be used to relate $\Delta V^0$ to $\Delta V^i$:
\begin{equation}\label{equ:k0}
    \sqrt{D^2+\left(\frac{l}{2}\right)^2}\Delta V^0=\frac{l}{2}\Delta V^1+D\Delta V^2
\end{equation}
The arrival time at dish B or dish A (with a reversing sign of $l$) is given by the integral $t_B=\int_0^1 dsV^0(s)$ and the difference to the lowest order in $l/D$ gives 
\begin{equation}\label{equ:delta_t}
    \Delta t=t_B-t_A=-\frac{h_+}{\omega_g}\cos(\omega_g t+\varphi_0)\sin\left(\frac{1}{2}\omega_g l\right)
\end{equation}
Here $t=D+t_0$ is the time of observation at the detectors. From the metric \eqref{equ:metric_pert}, we can see that the coordinate time equals the proper time of detectors, so $\eqref{equ:delta_t}$ directly gives the physical time difference we will measure. The $(\omega_gt+\varphi_0)$ term should be considered as the phase of local gravitational waves at the detector, which causes the oscillation of the arrival time difference between Dish A and Dish B. As shown in Eq.~\ref{equ:delta_t}, the arrival time difference only depends on the gravitational waves near the detector but not near the FRB source. 
As shown in \cite{Damour:1998jm}, the light deflection and arrival times by gravitational waves have no propagation effect but only depend on boundary terms on the source and observer.
Since FRB at different sightlines has not been separated yet at the source, the net effect on the arrival time difference induced by gravitational waves near the source is also zero. Therefore, the local GW is the only relevant boundary term that affects the arrival time differences among different detectors.
The oscillatory features of the arrival time difference described in Eq.~\ref{equ:delta_t} will be unique signatures for gravitational waves, which differ from the Shapiro time delays induced by dark matter halos \cite{Xiao:2024qay} or the expansion of the Universe \cite{Boone:2022pdz} that are either constant over time or a linear function of time. One possible contribution to the systematics comes from the Shapiro time delays of dark matter substructures with a size of our baseline (tens of AUs), which is roughly the size of QCD axion miniclusters \cite{Xiao:2021nkb,Xiao:2024qay}. Such objects can also induce large temporal changes in the arrival time difference but no periodicity is expected (On the other hand, GW signals will be important systematics if one wish to detect axion miniclusters).
The relevant GW frequency range in FRB timing is naturally determined by the baseline and the FRB repetition rate, which allows us to probe $\mu$Hz frequencies or smaller. The above expression has assumed the geometry depicted in Fig.~\ref{fig:setup} while generic FRB positions will introduce order unity corrections with the physical effect remaining the same. See Appendix.~\ref{app:general_deltat} for more details.

\section{Correlation Function of Signals}
One measurement of the FRB event will not achieve the full sensitivity of FRB timing, as the measurements for different sources can be correlated since they only depend on the phase of local gravitational waves up to some geometric factor.
Effective measurements of FRB timing give data as $(t_i,\Delta t_i)$, where $t_i$ is the time of measurement at either detector and $\Delta t_i$ is the arrival time difference. The angular positions of FRBs will also be needed for determining the order-unity geometric factors, which can be accurately measured from earth scale interferometry on FRBs \cite{Cassanelli:2021oaw}. Later, we will discuss the autocorrelation and angular correlation of arrival time difference, which can be used to further improve the sensitivity of FRB timing on GWs.

\subsection{Temporal Power Spectrum}
The arrival time difference induced by gravitational waves can be correlated at different times, and the temporal power spectrum can be calculated as
\begin{equation}\label{eq:temp_power}
    P(\omega)=\int\d \tau e^{-i\omega \tau}\left<\Delta t(\tau+t^{\prime})\Delta t(t^\prime)\right>_{t^\prime},
\end{equation}
where we have treated the arrival time difference $\Delta t$ as a function of arrival time and correlated them at different arrival times. Here the average $\left<\cdots\right>$ is taken over measurements at different $t^\prime$. Note that the autocorrelation function in Eq.~\ref{eq:temp_power} only depends on the difference in the measurement time, $\tau$. 
Consider the result of Eq.~\ref{equ:delta_t}, the correlation function of arrival time difference is directly given by the correlation function of metric perturbations induced by GW background, which can be further related to the spectral density of GWs, $S_h(\omega)$, and gives
\begin{equation}\label{eq:temp_signal}
P_t(\omega)=\frac{S_h(\omega)}{4\omega^2}F(\omega l),
\end{equation}
where the function $F(\Delta)$ is defined as
\begin{equation}\label{eq:FDelta}
    F(\Delta)=\frac{4}{3}+\frac{1}{\Delta^2}\cos\Delta-\frac{1+\Delta^2}{\Delta^3}\sin\Delta
\end{equation}
Therefore, the direct determination of the temporal power spectrum of the arrival time difference is also a direct measurement of the GW spectral density.

Our measurements of the arrival time difference are performed at discrete times, with a series of data $(t_i,\Delta t_i)$.
The temporal power spectrum can be obtained by performing the Fourier transform in discrete times.  The error of individual measurement is $\delta t_m$, and the number of repeating events is $N$, which will bring the variance of the timing error down to $\delta t_m^2/N$. Assuming the systematic uncertainties $\delta t_m$ from different measurements are independent, one could approximately treat the error on the autocorrelation function $\left<\Delta t(t+t^{\prime})\Delta t(t^\prime)\right>_{t^\prime}$ as the variance of the timing error.
Given the integration time, $2\pi/\omega$, for a given frequency of the Fourier mode, the error on the temporal power spectrum can be estimated as
\begin{equation}
    \delta P_t(\omega)\approx \frac{2\pi}{\omega} \frac{ \delta t_m^2}{N },\;\; {\rm for}\; \omega<\omega_0.
\end{equation}
Where $\omega_0=2\pi N/{T}$ is the highest frequency determined by the FRB repeating period ($N$ is the number of repeated events for one source), $\delta t_m=$ 0.1 ns is the timing precision of individual measurements, $T = 10$ yr is the total observation time. The above estimation is more accurate when the repeating of FRBs is periodic. Our above calculation on the uncertainty only serves for order-of-magnitude estimation of the sensitivity.
We have focused on the frequency range $2\pi/T<\omega<2\pi N/T$ for the calculation of GW temporal power spectrum, which is roughly the frequency band where the Fourier transform is sensible. For lower frequencies ($\omega<2\pi/T$), one could look for the GW signal as the time derivatives on the arrival time difference, as discussed in Appendix.~\ref{app:derivatives}. In this work, we focus on the statistical errors given by the number of measurements and the timing precision of individual measurements. There might be possible systematics, which has been assumed to not exceed the measurement error. The dominant systematics is likely the scattering effect in the interstellar medium of the Milky Way that leads to extra contribution to the time delay and thus the decoherence of FRBs, affecting the timing precision. However, the scattering effect is highly suppressed at higher frequencies and can be avoided with an observation frequency $f\gtrsim 5$ GHz \cite{Boone:2022pdz}. A better understanding of other noise sources will be needed to solidify our projected sensitivity level, which we leave for future explorations.

\begin{figure}
    \centering
    \includegraphics[width=\linewidth]{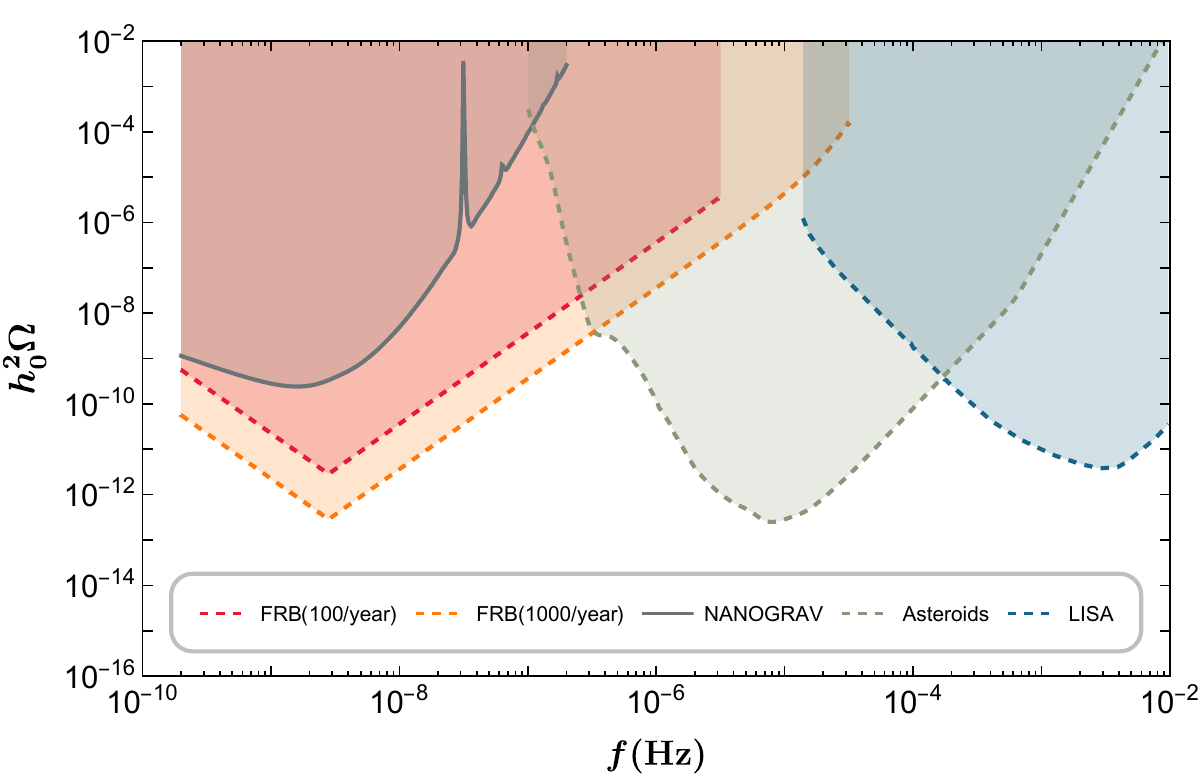}
    \caption{
    In this plot, we present our projected sensitivity of FRB timing as a function of GW frequency and compare it to other existing or proposed observations. The y-axis represents the energy fraction in the Universe contributed by gravitational waves ($\Omega_{\rm gw}$) multiplied by $h_0^2$ with $h_0=0.67$ gives the current Hubble \cite{Planck:2018vyg}.
    The red and orange dashed curves show the sensitivity projections by using the temporal power spectrum and other temporal changes of the arrival time difference, assuming a dish separation of 50AU, 100 or 1000 FRB events detected per year and an observation time of 10 years. The upper bound of frequency is determined by the detection rate of FRB. 
    Our sensitivity at the higher frequency end is obtained by measuring the temporal power spectrum of GW signals. At lower frequencies, this method breaks down, and we switch to looking at the second derivatives of the arrival time difference to constrain GW-induced time delays, which causes the break on the sensitivity curve. Our timing precision of individual FRB events is 0.1 ns and it is assumed our measurements are not affected by other systematics. See the main text for more discussions.
    The grey solid curve is from the NANOGRAV sensitivity \cite{NANOGrav:2023ctt}, the blue dashed curve is from LISA projections \cite{amaro2017laser}, and the green dashed curve is from asteroids projection \cite{Fedderke:2021kuy}. 
    }
    \label{fig:fullnomodel}
\end{figure}

The sub-nanosecond timing precision roughly corresponds to $1/f$ where $f$ is the observation frequency of FRBs ($\gtrsim$ GHz), which is achieved through correlating the electric field series of the burst profile. 
The previous expression of the temporal power spectrum has assumed a special geometry, i.e. the FRB source is located on the perpendicular bisector of the line connecting the two dishes as shown in Fig.~\ref{fig:setup}. In reality, FRB sources should be randomly distributed in every direction and the signals from different sources are still correlated since the arrival time difference only depends on the phase of local gravitational waves. Averaging over all directions from all FRB sources, we can obtain the expected temporal power spectrum and thus the following sensitivity on the fraction of total energy density in the universe contributed by GWs
\begin{equation}\label{equ:auto}
  \delta\Omega_{\mathrm{gw}}= \frac{4\omega^{4}\delta t_m^2}{3NN_{\mathrm{FRB}}H_0^2}\left[F_{\mathrm{auto}}(\omega l)\right]^{-1},
\end{equation}
where $N$ is the number of repeating events for each FRB source and $N_{\rm FRB}$ is the number of sources that are actively repeating. Therefore $N_{\rm burst}=N N_{\rm FRB}$, the total number of bursts detected, is the only relevant parameter for our sensitivity. The best strategy to maximize the number of bursts detected per unit time might involve targeting active phases of FRB repeaters and monitor those bright sources with a high repeating rate.
The function $F_{\mathrm{auto}}$ that factorize the frequency dependence is defined by
\begin{align}
    F_{\mathrm{auto}}(\Delta)=&\frac{4\Delta^6-6\Delta^4-6\Delta^2-9}{3\Delta^6}\nonumber\\
    &+\frac{3-4\Delta^2}{\Delta^6}\cos(2\Delta)+\frac{6}{\Delta^5}\sin(2\Delta)
\end{align}
See more detailed derivations in Appendix \ref{sec:corr}. In the frequency band of interest, the parameter $\Delta$ satisfies $\Delta=\omega l\leq 0.3$. Therefore, we may approximate $F_{\mathrm{auto}}(\Delta)$ by the first order of its Taylor expansion,

\begin{equation}
    F_{\mathrm{auto}}(\Delta)= \frac{2}{9}\Delta^2+\mathcal{O}(\Delta^4)
\end{equation}
So overall, the sensitivity on gravitational waves goes like $\delta\Omega_{\mathrm{gw}}\propto \omega^{2}l^{-2}$, indicating that a longer baseline results in a better sensitivity. The sensitivity obtained in Eq.~\ref{equ:auto} will apply to frequencies $\omega\gtrsim 2\pi/T$. When we switch to lower frequencies, we can use the time derivatives of the arrival time difference to place constraints on the GW signal (see Appendix.~\ref{app:derivatives}), which will cause a break in the power law sensitivity curves, as shown in Fig.~\ref{fig:fullnomodel}.

Assuming a baseline of 50 AU, a observation time of 10 years, and a detection rate of 100 FRB events per year, we find that our proposal fills the gap between $0.1\mu$Hz and $1\mu$Hz, as shown in Fig.~\ref{fig:fullnomodel}, and our projected sensitivity is competitive in the nHz frequency range and complementary to the asteroid proposal that has a better sensitivity at higher frequencies. More sophisticated statistical methods are needed in the future for realistic estimations of measurement errors and sensitivities.
A larger number of FRB events observed by the detectors or a longer baseline can lead to greater sensitivity. Here we take the total number of FRB events per year to be 100 to 1000, which is on the reach of a 10-meter dish in space (see estimations of FRB observation rate in Ref.~\cite{Boone:2022pdz,Xiao:2022hkl}). Radio telescopes on Earth can assist in the search for active FRB repeaters and the space dishes can point to known bright FRB repeaters, which will greatly enhance the rate of detecting FRB events.

\subsection{Angular Correlation}
Similar to the derivation of the Hellings Downs curve \cite{Hellings:1983fr,Romano:2023zhb}, we can correlate the arrival time difference induced by GWs for FRBs at different angular positions. 
We view $\Delta t$ as a function of detection time and the two angular coordinates $\theta_0$, $\phi_0$, i.e.
\begin{equation}
    \Delta t=\Delta t(t,\theta_0,\phi_0)
\end{equation}
Here $\theta_0$, $\phi_0$ are the usual spherical coordinates, with $\theta_0$ being the angle between the FRB and the line connecting the two dishes, and $\phi_0$ being the angle of rotation around that line. The correlation function can be expressed as
\begin{equation}
    C(t,\theta_1,\theta_2,\phi_2-\phi_1)=\left<\Delta t(t,\theta_1,\phi_1)\Delta t(0,\theta_2,\phi_2)\right>
\end{equation}
The average should be taken over different FRB events measured at different times, with the time difference expected to be within one cycle of gravitational waves.
For simplicity, we fix the position of one FRB source to be $\theta_1=0$, which can be achieved when there are enough sources observed by our dishes. Thus, we extract the angular dependence of the correlation function and denote it by $\alpha_{\mathrm{angular}}$, 
\begin{equation}
    C_{\mathrm{angular}}(t,\theta_2)=\int_{-\infty}^\infty\d f e^{i2\pi ft} \frac{S_h(f)}{32\pi^2 f^2}\alpha_{\mathrm{angular}}(\Delta,\theta_2)
\end{equation}

Here $\Delta$ is again defined as $\Delta=2\pi f\,l$ with $l$ being the dish separation. The analytic form of $\alpha_{\mathrm{angular}}$ is rather complicated. Fortunately, the parameter $\Delta$ in the frequency range of interest satisfies $\Delta\lesssim 1$ and $\alpha_{\mathrm{angular}}$ can also be expanded in powers of $\Delta$, which gives
\begin{equation}
    \alpha_{\mathrm{angular}}(\Delta,\theta_2)=\frac{\Delta^2}{30}(11\cos\theta_2+5\cos3\theta_2)+\mathcal{O}(\Delta^4),
\end{equation}
where higher order terms $\mathcal{O}(\Delta^4)$ can be found in Appendix \ref{sec:corr} for more accurate determination of the angular correlation function. The angular correlation function we obtained for FRB timing differs from that in the Hellings Downs curve \cite{Hellings:1983fr,Romano:2023zhb} because we have two detectors and the positioning of two dishes and the source is relevant for the angular correlation while pulsar timing only needs one detector. If measured in the future, the angular correlation function of FRB timing can be another piece of evidence to confirm the detection of gravitational waves.


\section{Discussion}
We use the FRB timing technique to measure local gravitational waves in the frequency range of nHz to $\mu$Hz, which can fill the `$\mu$Hz' gap in GW detection. While FRBs are not clocks in themselves, observing the same FRB source among different sightlines can achieve subnanosecond timing precision on the arrival time difference through coherent analysis of FRB signal received at different detectors. Therefore, the reference clock in FRB timing can be realized by sending radio telescopes to space and observing the same FRB source with different detectors. The measurement of the arrival time difference at different detectors will be sensitive to local gravitational waves. Since we measure the difference between detectors, the GW contribution near the FRB source will not affect the measurement. Therefore, every measurement of FRB arrival time difference induced by GWs can be correlated both in time and space. As a result, the sensitivity of FRB timing can be improved significantly if we monitor known FRB sources that are actively repeating to increase the detection rate of bursts. A conservative estimation based on a FRB detection rate of 100 events per year already gives a sensitivity that significantly improves that of pulsar timing and extends the reach of GW to higher frequencies.

This new method requires ambitious space missions to achieve solar-system scale interferometry on FRBs, which has already been proposed to measure cosmological distances to sub-percent precision and detect dark matter substructures on extremely small scales. Our theoretical calculation on the signal of gravitational waves will further increase the scientific opportunities of this potential mission. Understanding the systematics is crucial to achieving the desired timing accuracy in all proposals using FRB timing. The primary systematics is likely the scattering effect in the interstellar medium of the Milky Way that causes the decoherence of FRBs, which affects the timing precision. However, the scattering effect rapidly goes down at higher frequencies so this constraint can be avoided when observing FRBs at frequencies $f\gtrsim 5$ GHz \cite{Boone:2022pdz}. Although extensive studies on the systematics are necessary for a realistic prediction of the sensitivity, the current findings are promising and we expect FRB timing to be a competitive proposal for the detection of GWs in the frequency range of nHz to $\mu$Hz.

\section*{Acknowledgement}
We thank Liang Dai, Matthew Mcquinn, Kai Schmitz for helpful discussions and suggestions on the draft.
HX is supported by Fermi Research Alliance, LLC under Contract DE-AC02-07CH11359 with the U.S. Department of Energy. The work of L.T.W. is supported by DOE grant
DE-SC-0013642. 
\onecolumngrid
\appendix

\section{Metric perturbations from Stochastic Gravitational Waves}\label{sec: metric}
In this section we discuss the modeling of stochastic gravitational waves, our convention follows \cite{Maggiore:2007ulw}. Assume there is some gravitation wave background $h_{ab}(\bs{x},t)$ in our universe, this can always be expanded into a combination of monochromatic modes, 
\begin{equation}\label{equ:metric_expansion}
    h_{ab}(\bs{x},t)=\int_{-\infty}^{\infty}\d f\int \d^2\Omega \tilde{h}_{ab}(f,\hat{n})e^{i2\pi f(t-\hat{n}\cdot \bs{x})}
\end{equation}
Here the indices $a,b$ runs from 1 to 3. The negative frequency modes satisfies $\tilde{h}_{ab}(-f,\hat{n})=\tilde{h}_{ab}^*(f,\hat{n})$. Here we have adopted the TT gauge, i.e. 
\begin{equation}
    h_{00}=h_{0a}=h_{aa}=\partial_{a}h_{ab}=0
\end{equation}
The metric perturbations split into two polarizations,
\begin{equation}\label{app:equ:modes}
    \tilde{h}_{ab}(f,\hat{n})=\tilde{h}_{+}(f,\hat{n})e_{ab}^+(\hat{n})+\tilde{h}_{\times}(f,\hat{n})e_{ab}^\times(\hat{n})
\end{equation}
Introduce two unit vectors $\hat{l}$ and $\hat{m}$, such that $\hat{l}$, $\hat{m}$ and $\hat{n}$ are orthogonal to each other. Particularly, one may choose $\hat{l}=\hat{\theta}$ and $\hat{m}=\hat{\phi}$ in spherical coordinates . The two polarization matrices can be expressed as 
\begin{equation}\label{app:equ:e+}
    e_{ab}^+(\hat{n})=\hat{l}_a\hat{l}_b-\hat{m}_a\hat{m}_b
\end{equation}
\begin{equation}\label{app:equ:ecross}
    e_{ab}^\times(\hat{n})=\hat{l}_a\hat{m}_b+\hat{m}_a\hat{l}_b
\end{equation}
It's easy to verify that this definition is dependent on the choice of $\hat{l}$ and $\hat{m}$. The polarization tensors satisfy an orthonormal condition, 
\begin{equation}
    \sum_{a,b}e_{ab}^{A}e_{ab}^{A'}=2\delta^{A A'}
\end{equation}
where $A$ and $A'$ stands for $+$ or $\times$.
The energy density of gravitational waves can be written as the power of metric perturbations: 
\begin{equation}\label{equ:rho}
    \rho_{gw}=\frac{1}{32\pi G}\left<\sum_{a,b}\dot{h}_{ab}(t^\prime)\dot{h}_{ab}(t^\prime)\right>
\end{equation}
Here the average is taken over ensembles. In practice, there is only one physical system, which is our detector, but we can keep measuring the signal over time, and the ensemble average can be replaced by a time average. We model the stochastic GW background in the following way:
\begin{equation}\label{app:equ:expctation}
   \left<\tilde{h}_{A}(f,\hat{n})\tilde{h}^*_{A'}(f',\hat{n}')\right>=\frac{1}{8\pi}\delta(f-f')\delta^2(\hat{n},\hat{n}')\delta_{AA'}S_h(f)
\end{equation}
Here $A$ and $A'$ stand for $+$ or $\times$, as introduced in Eq.~\ref{app:equ:modes}. $S_h(f)$ is called the spectral density of the stochastic background, it satisfies $S_h(-f)=S_h(f)$ and has a dimension of inverse frequency. We have assumed that the gravitational wave background is isotropic, so the spectral density $S_h$ is independent of $\hat{n}$. The delta function on the angular part can be explicitly written as
\begin{equation}
    \delta^2(\hat{n},\hat{n}')=\delta(\phi-\phi')\delta(\cos\theta-\cos\theta')
\end{equation}
Insert it into \eqref{equ:rho}, one can obtain the following expression for the energy density of gravitational waves
\begin{equation}
    \rho_{gw}=\frac{\pi}{4 G}\int_{-\infty}^{\infty}\d f f^2 S_h(f)=\frac{\pi}{2 G}\int_{0}^{\infty}\d f f^2 S_h(f)
\end{equation}

The fractional energy spectrum is defined by the energy density of GW per logarithmic frequency bin normalized by the critical density of the Universe, which gives
\begin{equation}
    \Omega_{gw}(f)=\frac{f}{\rho_{\mathrm{cric}}}\frac{\d\rho_{gw}}{\d f}
\end{equation}
Here the critical density is simply
\begin{equation}
    \rho_{\mathrm{cric}}=\frac{3H_0^2}{8\pi G},
\end{equation}
where $H_0$ is the Hubble constant. Putting everything together, we have
\begin{equation}
    \Omega_{gw}(f)=\frac{4\pi^2}{3H_0^2}f^3S_h(f)
\end{equation}
The fractional energy spectrum is related to the characteristic strain by
\begin{equation}
    \Omega_{gw}(f)=\frac{2\pi^2}{3H_0^2}f^2h_c^2(f)
\end{equation}
Therefore,
\begin{equation}
    h_c^2(f)=2fS_h(f)
\end{equation}
The characteristic strain $h_c(f)$ is dimensionless. 
The Hubble constant can be written in the form
\begin{equation}
    H_0=h_0\times 100\mbox{km s}^{-1}\mbox{ Mpc}^{-1}
\end{equation}
There are several physical quantities that characterize the intensity of stochastic gravitational wave background, such as the strain $h_c(f)$, spectral density $S_h(f)$, and $h_0^2\Omega_{gw}(f)$. One can translate them using the above expressions.

\section{General Treatments on the Arrival Time Difference}\label{app:general_deltat}
In the main text, we have computed the arrival time difference caused by a monochromatic plane gravitational wave in a special geometry setup. In reality, however, the FRB sources are located at random angular positions, and the gravitational wave background is composed of plane waves with all possible frequencies and propagation directions. In this section, we will calculate the arrival time difference with the most general setup. 
\begin{figure}
    \centering
    \includegraphics[width=0.5\linewidth]{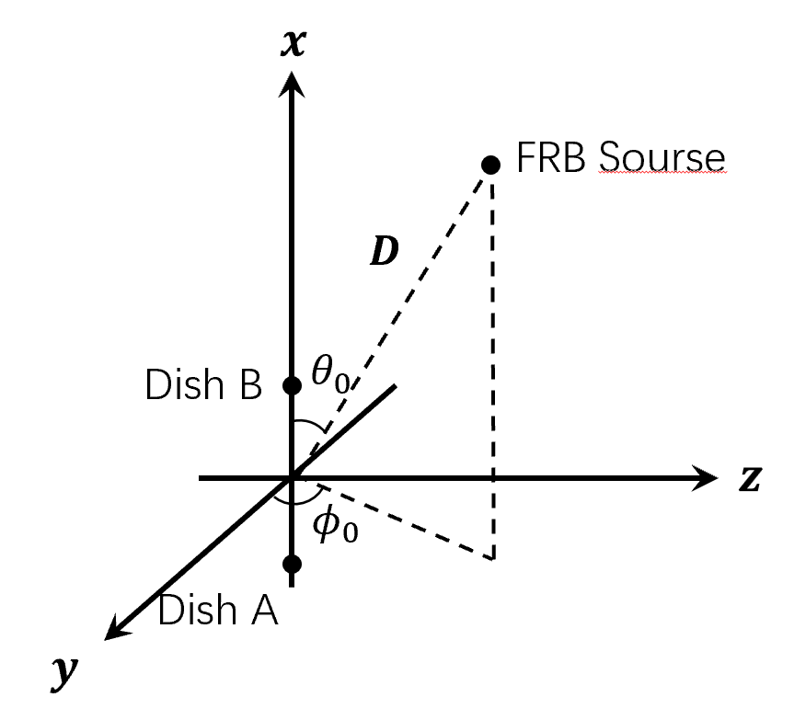}
    \caption{Geometric setup. We construct the coordinate system so that the two detectors are on the $x$ axis. The FRB source is located at an arbitrary position in the given coordinate system. We use $\theta_0$ and $\phi_0$(as shown in the figure) to describe the position of the FRB source. The distance between the source and the origin is $D$ ($\sim $Gpc). }
    \label{fig:general_setup}
\end{figure}
The geometry is shown in Fig. \ref{fig:general_setup}, we have two detectors on the $x$ axis and an FRB source at an arbitrary position described by $\theta_0$ and $\phi_0$. The distance between the source and the origin is denoted by $D$. 
We will consider a metric perturbation as in \eqref{equ:metric_expansion}. The photon path is described by a set of four functions,
\begin{equation}
    x^\mu(s)=(t(s),x(s),y(s),z(s))
\end{equation}
with $s$ being the affine parameter, taking its value in $[0,1]$. $s=0$ corresponds to the emission event, and $s=1$ corresponds to the absorption event. The four velocity is defined by the derivative of coordinates over $s$
\begin{equation}
    V^\mu=\frac{\d x^\mu}{\d s}
\end{equation}

First let's consider the situation without gravitational waves. We add a bar to the physical quantities before perturbation. Now the coordinate functions of the photon path are denoted by
\begin{equation}
    \bar{x}^\mu(s)=(\bar{t}(s),\bar{x}(s),\bar{y}(s),\bar{z}(s))
\end{equation}
Correspondingly, the four velocity is
\begin{equation}
    \bar{V}^\mu=\frac{\d \bar{x}^\mu}{\d s}
\end{equation}
Take dish B for example, without gravitational waves, light travels in a straight line, so the unperturbed four velocity is
\begin{align}
    &\bar{V}^0=\sqrt{D^2+\left(\frac{l}{2}\right)^2-Dl\cos\theta_0},\qquad \bar{V}^1=\frac{l}{2}-D\cos\theta_0,\nonumber\\ 
    &\bar{V}^2=-D\sin\theta_0\cos\phi_0,\qquad \bar{V}^3=-D\sin\theta_0\sin\phi_0
\end{align}
Accordingly, the coordinate functions are
\begin{equation}
    \bar{x}^\mu=\bar{V}^\mu s
\end{equation}
The FRB emission event is at
\begin{equation}
    x_0^\mu=(t_0,D\cos\theta_0,D\sin\theta_0\cos\phi_0,D\sin\theta_0\sin\phi_0)
\end{equation}
For a simpler notation, we define a three-dimensional unit vector 
\begin{equation}
    \hat{s}=(\cos\theta_0,\sin\theta_0\cos\phi_0,\sin\theta_0\sin\phi_0),
\end{equation}
which is the direction of the FRB source.
When we add the gravitational wave background to the system, the coordinate functions should change by a term proportional to the perturbation $h$, namely $\tilde{x}^\mu(s)=x^\mu(s)+\mathcal{O}(h)$. We know that in TT gauge, the coordinates of test masses stay fixed, so we have boundary conditions for the coordinate functions,
\begin{equation}\label{equ:bound}
    x^\mu(0)=x^\mu_0=\bar{x}^\mu(0),\ x^a(1)=\mbox{spatial position of detector}=\bar{x}^a(1)
\end{equation}
After adding the perturbation, the four-velocity should also change by a term proportional to $h$, which can be written as
\begin{equation}
    \bar{V}^\mu+u^\mu(s)=V^\mu=\frac{\d x^\mu}{\d s}
\end{equation}
This four velocity satisfies the geodesic equation, so to the first order in $h$, we have a set of equations for $u^\mu(s)$,
\begin{equation}\label{equ:gen_time}
    \frac{\d u^0}{\d s}+\frac{1}{2}\bar{V}^a\bar{V}^b\partial_0h_{ab}(\bar{V}s+x_0)=0
\end{equation}
\begin{equation}\label{equ:gen_space}
    \frac{\d u^a}{\d s}+\bar{V}^b\bar{V}^\mu\partial_\mu h_{ab}(\bar{V}s+x_0)-\frac{1}{2}\bar{V}^b\bar{V}^c\partial_ah_{bc}(\bar{V}s+x_0)=0
\end{equation}
To simplify the notation, we will denote the measure collectively by $\d \Phi$, i.e.
\begin{equation}
    \int_{-\infty}^{\infty}\d f\int \d^2\Omega \Rightarrow \int \d\Phi
\end{equation}
Also, we will denote the GW wave vector by
\begin{equation}
    k_g^\mu=2\pi f(1,\hat{n})
\end{equation}
Now we insert the Fourier transformation \eqref{equ:metric_expansion} and the geodesic equations become
\begin{equation}
    \frac{\d u^0}{\d s}+\frac{i}{2}\int\d\Phi e^{-i\bar{V}\cdot k_g s-ik_g\cdot x_0} k_g^0 \bar{V}^a \bar{V}^b\tilde{h}_{ab}=0
\end{equation}
\begin{equation}
    \frac{\d u^a}{\d s}-i\int\d\Phi e^{-i\bar{V}\cdot k_g s-ik_g\cdot x_0} \bar{V}\cdot k_g \bar{V}^b \tilde{h}_{ab}+\frac{i}{2}\int\d\Phi e^{-i\bar{V}\cdot k_g s-ik_g\cdot x_0} \bar{V}^b\bar{V}^c k_g^a \tilde{h}_{bc}=0
\end{equation}
One can easily integrate the two functions over $s$, now
\begin{equation}
    u^0=\Delta u^0+\int\d\Phi e^{-i\bar{V}\cdot k_g s-ik_g\cdot x_0} \frac{k_g^0}{2 \bar{V}\cdot k_g} \bar{V}^a \bar{V}^b\tilde{h}_{ab}
\end{equation}
\begin{equation}
    u^a=\Delta u^a + \int\d\Phi e^{-i\bar{V}\cdot k_g s-ik_g\cdot x_0} \left(-\bar{V}^b \tilde{h}_{ab}+\frac{k_g^a}{2 \bar{V}\cdot k_g}\bar{V}^b \bar{V}^c\tilde{h}_{bc}\right)
\end{equation}
$\Delta u^\mu$ is a set of four integration constants. At the four velocity level, we know that the four velocity of a photon always have zero norm, this puts a constraint on $\Delta u^\mu$. By direct calculation, one gets
\begin{equation}
    \bar{V}^0 \Delta u^0=\bar{V}^a \Delta u^a
\end{equation}
To completely determine $\Delta u^\mu$, one has to use the boundary condition \eqref{equ:bound}. To do this, integrate the four velocity over $s$ from 0 to 1. Written explicitly, one has another set of equations 
\begin{equation}
    t=\bar{V}^0+\Delta u^0+\int\d\Phi e^{-ik_g\cdot x_0}(e^{-i\bar{V}\cdot k_g}-1) \frac{ik_g^0}{2 (\bar{V}\cdot k_g)^2} \bar{V}^a \bar{V}^b\tilde{h}_{ab}
\end{equation}
\begin{equation}
    0=\Delta u^a+\int\d\Phi \frac{ie^{-ik_g\cdot x_0}(e^{-i\bar{V}\cdot k_g}-1)}{\bar{V}\cdot k_g} \left(-\bar{V}^b \tilde{h}_{ab}+\frac{k_g^a}{2 \bar{V}\cdot k_g}\bar{V}^b \bar{V}^c\tilde{h}_{bc}\right)
\end{equation}
Putting everything together, we have the arrival time
\begin{equation}
    t_B\equiv \tilde{t}(1)=t_0+\bar{V}^0+\int\d\Phi e^{-ik_g\cdot x_0}(e^{-i\bar{V}\cdot k_g}-1)\frac{i\bar{V}^a \bar{V}^b\tilde{h}_{ab}}{2\bar{V}^0 \bar{V}\cdot k_g}
\end{equation}
There is a similar expression for $t_A$, one only has to change $l\rightarrow -l$.

$l$ is on the scale of $100$ AU, and $D$ is on the scale of 1 Gpc, so we may keep only the lower order in $l/D$. The time difference is
\begin{equation}\label{app:equ:dt}
    \Delta t=t_B-t_A=-l\cos\theta_0-\int\d\Phi e^{i2\pi f(t_0+D)}\sin\left[\pi fl\left(\cos\theta_0+n_x\right)\right]\frac{\hat{s}_a\hat{s}_b\tilde{h}_{ab}}{2\pi f(1+\hat{n}\cdot\hat{s})}
\end{equation}
The first term is purely geometric, and it doesn't change with time. This geometric term should be already subtracted from the data. Even if it is not subtracted, this term would only cause a sharp peak at $f=0$ in the temporal power spectrum, so in either case, this term has no effect on the final result. From now on, we will omit this geometric term in the derivation. 

\section{General Treatments on the Correlation Function}\label{sec:corr}
In this section, we compute the correlation function of the signals. Similar to the Hellings Downs curve, the way that the correlation function depends on angular coordinates reflects the nature of the metric perturbation. By measuring the depedency of the correlation function over angular coordinates, we can distinguish gravitational wave from other perturbations that could cause a arrival time difference.
As shown in the last section, $\Delta t$ is a function of $t=t_0+D$ and the two angular coordinates $\theta_0$, $\phi_0$, i.e.
\begin{equation}
    \Delta t=\Delta t(t,\theta_0,\phi_0)
\end{equation}
The correlation function is defined by
\begin{equation}\label{app:equ:cordef}
    C(t,\theta_1,\theta_2,\phi_2-\phi_1)=\left<\Delta t(t,\theta_1,\phi_1)\Delta t(0,\theta_2,\phi_2)\right>
\end{equation}
Note that since we have a dipole detector, the two detectors determine a special direction in space, so generically we need three angular coordinates to specify the correlation function. This is quite different from the Hellings Downs curve used in PTA, where only one detector is used so the correlation function depends only on the relative angle between two pulsars.

Insert Eq. \eqref{app:equ:dt} into Eq. \eqref{app:equ:cordef},
\begin{equation}
    C(t,\theta_1,\theta_2,\phi_2-\phi_1)=\int\d\Phi\d\Phi' e^{i2\pi ft}\sin\left[\pi fl\left(\cos\theta_1+n_x\right)\right]\sin\left[\pi f'l\left(\cos\theta_2+n'_x\right)\right]\frac{\hat{s}_{1a}\hat{s}_{1b}\hat{s}_{2c}\hat{s}_{2d}\left<\tilde{h}_{ab}(f,\hat{n})\tilde{h}_{cd}(f',\hat{n}')\right>}{4\pi^2 ff'(1+\hat{n}\cdot\hat{s}_1)(1+\hat{n}'\cdot\hat{s}_2)}
\end{equation}
Here for notational simplicity, we have adopted the Einstein summation convention, the indices $a,b,c,d$ are implicitly summed. From equations \eqref{app:equ:modes} and \eqref{app:equ:expctation}, we know
\begin{equation}
    \left<\tilde{h}_{ab}(f,\hat{n})\tilde{h}_{cd}(f',\hat{n}')\right>=\frac{S_h(f)}{8\pi}\delta(f+f')\delta^2(\hat{n},\hat{n}')(e_{ab}^+(\hat{n})e_{cd}^+(\hat{n})+e_{ab}^\times(\hat{n})e_{cd}^\times(\hat{n}))
\end{equation}
The integral over $\d\Phi'$ cancels with the two $\delta$ functions. Inserting equations \eqref{app:equ:e+} and \eqref{app:equ:ecross}, we get the final result,
\begin{align}\label{app:equ:correlation}
     & C(t,\theta_1,\theta_2,\phi_2-\phi_1)=\int\d\Phi e^{i2\pi ft}\frac{S_h(f)}{8\pi}\frac{\sin(\pi lf(\cos\theta_1+n_x))\sin(\pi lf(\cos\theta_2+n_x))}{(2\pi f)^2(1+\hat{n}\cdot \hat{s_1})(1+\hat{n}\cdot \hat{s_2})}\nonumber                          \\
     & \times \left\{\left[(\hat{s}_1\cdot\hat{l})^2-(\hat{s}_1\cdot\hat{m})^2\right]\left[(\hat{s}_2\cdot\hat{l})^2-(\hat{s}_2\cdot\hat{m})^2\right]+4(\hat{s}_1\cdot\hat{l})(\hat{s}_1\cdot\hat{m})(\hat{s}_2\cdot\hat{l})(\hat{s}_2\cdot\hat{m})\right\}
\end{align}
The integral cannot be carried out analytically. Here, we specify on two important cases.

\subsection*{Auto-correlation}

As mentioned before, the two detectors determine a special direction in space, so the auto-correlation of one FRB already contains one angular coordinate. Set $\phi_1=\phi_2$, $\theta_1=\theta_2$ in \eqref{app:equ:correlation}, we get
\begin{equation}
    C_{\mathrm{auto}}(t,\theta_1)=\int_{-\infty}^\infty\d f e^{i2\pi ft} \frac{S_h(f)}{32\pi^2 f^2}\alpha_{\mathrm{auto}}(\Delta,\theta_1)
\end{equation}
where
\begin{align}
    \alpha_{\mathrm{auto}}(\Delta,\theta_1)=
     & \left(\frac{8}{3}-\frac{1}{\Delta^2}(1+3\cos2\theta_1)\cos\Delta\cos(\Delta\cos\theta_1)-\frac{4}{\Delta^2}\cos\theta_1\sin\Delta\sin(\Delta\cos\theta_1)\right.\nonumber \\
     & \left.-\frac{-1+3\Delta^2+(\Delta^2-3)\cos2\theta_1}{\Delta^3}\sin\Delta\cos(\Delta\cos\theta_1)+\frac{4}{\Delta}\cos\theta_1\cos\Delta\sin(\Delta\cos\theta_1)\right)
\end{align}
\begin{figure}
    \centering
    \subfloat[$\Delta=1$]{\includegraphics[width=0.48\linewidth]{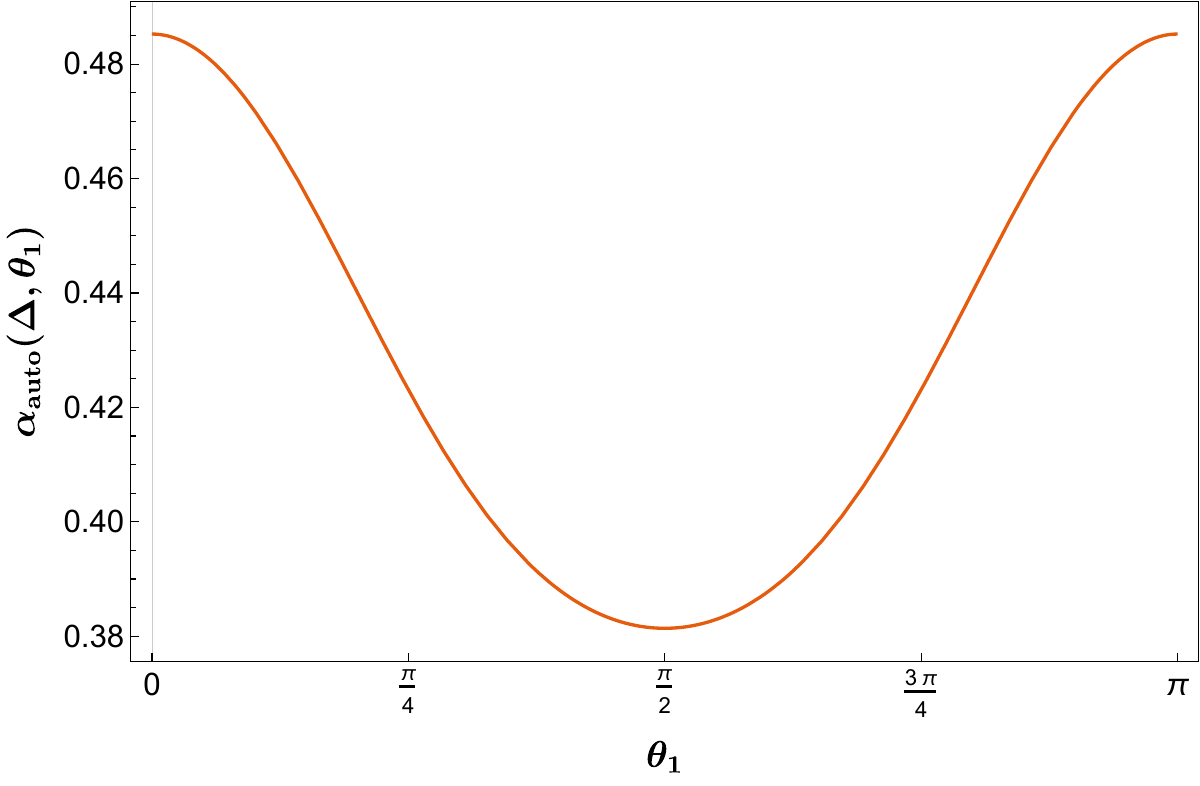}\label{app:fig:auto_d=1}}
    \subfloat[$\Delta=3$]{\includegraphics[width=0.48\linewidth]{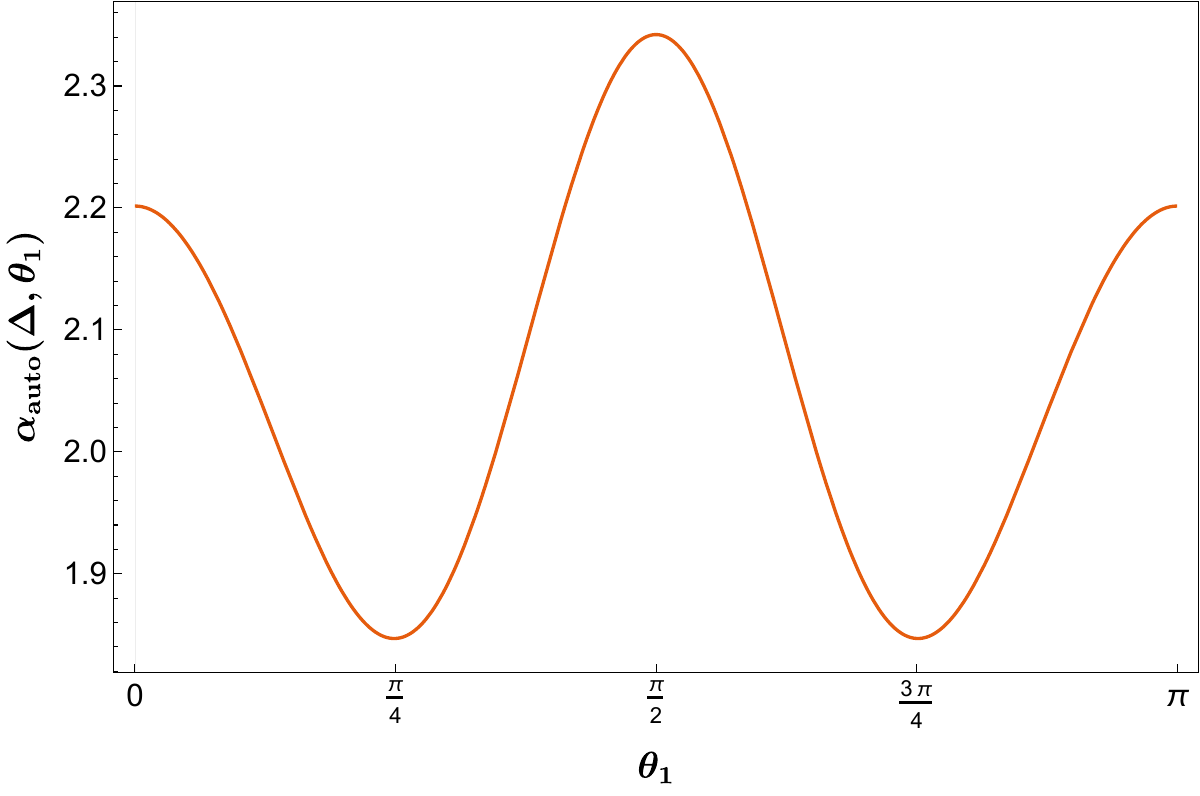}\label{app:fig:ang_d=3}}

    \subfloat[$\Delta=7$]{\includegraphics[width=0.48\linewidth]{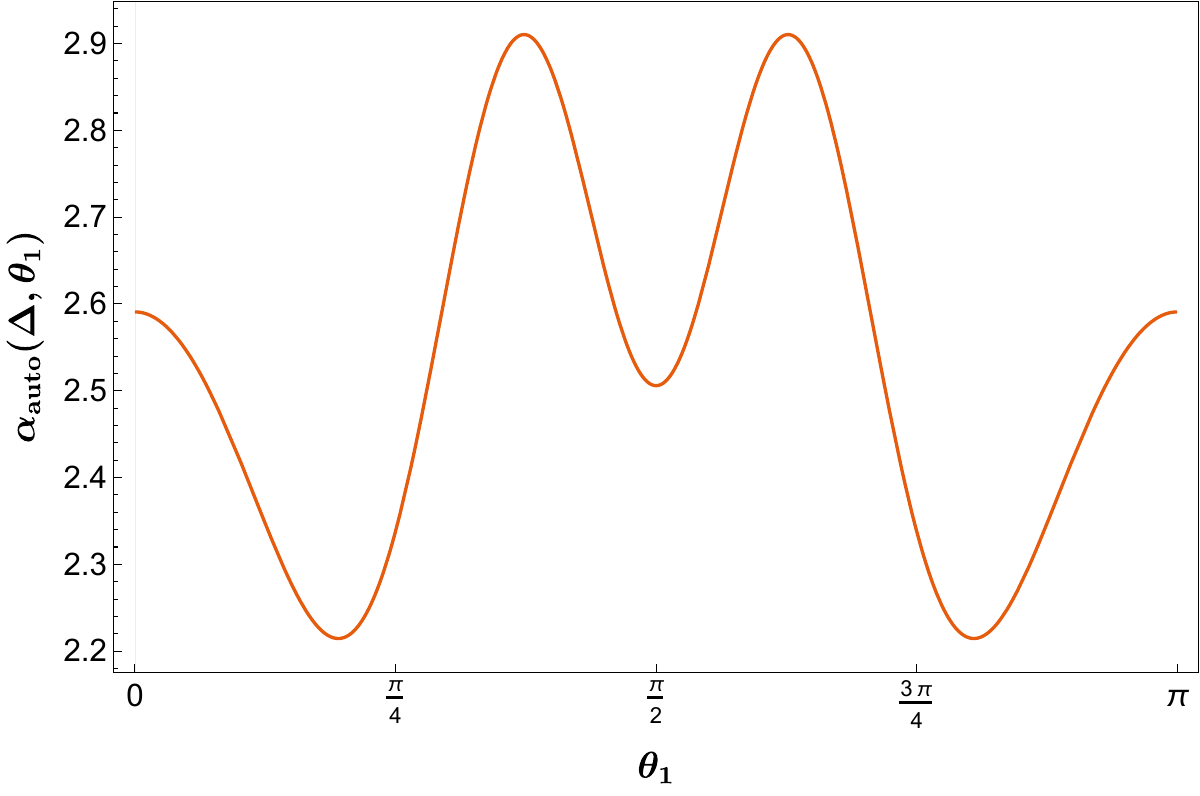}\label{app:fig:ang_d=7}}
    \subfloat[$\Delta=10$]{\includegraphics[width=0.48\linewidth]{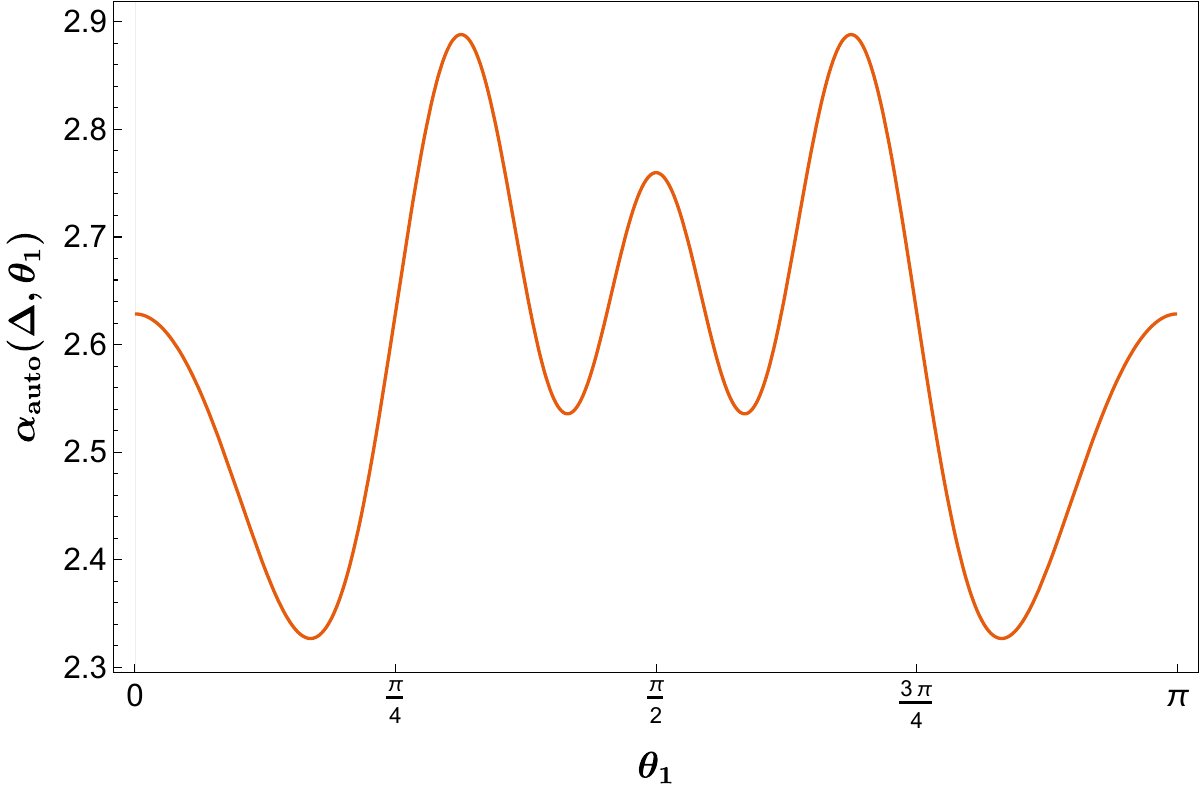}\label{app:fig:ang_d=10}}
    \caption{This figure presents the autocorrelation function, $\alpha_{\mathrm{auto}}(\Delta,\theta_1)$ as a function of the angular position of FRB source $\theta_1$ at different GW frequencies ( $\Delta=1,\ 3,\ 7,\ 10$). The oscillatory feature in these plots originates from the nature of GWs. }\label{app:fig:auto}
\end{figure}
Here $\Delta=2\pi fl=\omega_g l$. To get some intuition, we present here the plots of $\alpha_{\mathrm{auto}}(\Delta,\theta_1)$ over $\theta_1$ for four different $\Delta$, see Fig. \ref{app:fig:auto}.

Now let's calculate the sensitivity we can get by using auto-correlation. The auto-correlation of one FRB source would contribute to a signal in all frequency bins. We can add up the signal from all of the FRBs we can observe. In this process, the sensitivity by $N_{\mathrm{FRB}}$. The averaged power spectrum is
\begin{equation}
    \bar{P}_t(\omega)=\frac{S_h(\omega)}{4\omega^2}\left(\frac{4\Delta^6-6\Delta^4-6\Delta^2-9}{3\Delta^6}+\frac{3-4\Delta^2}{\Delta^6}\cos(2\Delta)+\frac{6}{\Delta^5}\sin(2\Delta)\right)
\end{equation}
The constraint would be
\begin{equation}\label{app:equ:autoreach}
    \Omega_{\mathrm{gw}}= \frac{4\omega^{4}\delta t_m^2}{3N N_{\mathrm{FRB}} H_0^2}\left(\frac{4\Delta^6-6\Delta^4-6\Delta^2-9}{3\Delta^6}+\frac{3-4\Delta^2}{\Delta^6}\cos(2\Delta)+\frac{6}{\Delta^5}\sin(2\Delta)\right)^{-1}
\end{equation}
Or using the characteristic strain $h_c$,
\begin{equation}\label{app:equ:autoreach_h}
    h_c= \left[\frac{8\omega^{2}\delta t_m^2}{NN_{\mathrm{FRB}}}\left(\frac{4\Delta^6-6\Delta^4-6\Delta^2-9}{3\Delta^6}+\frac{3-4\Delta^2}{\Delta^6}\cos(2\Delta)+\frac{6}{\Delta^5}\sin(2\Delta)\right)^{-1}\right]^{1/2}
\end{equation}
\begin{figure}
    \centering
    \includegraphics[width=0.8\linewidth]{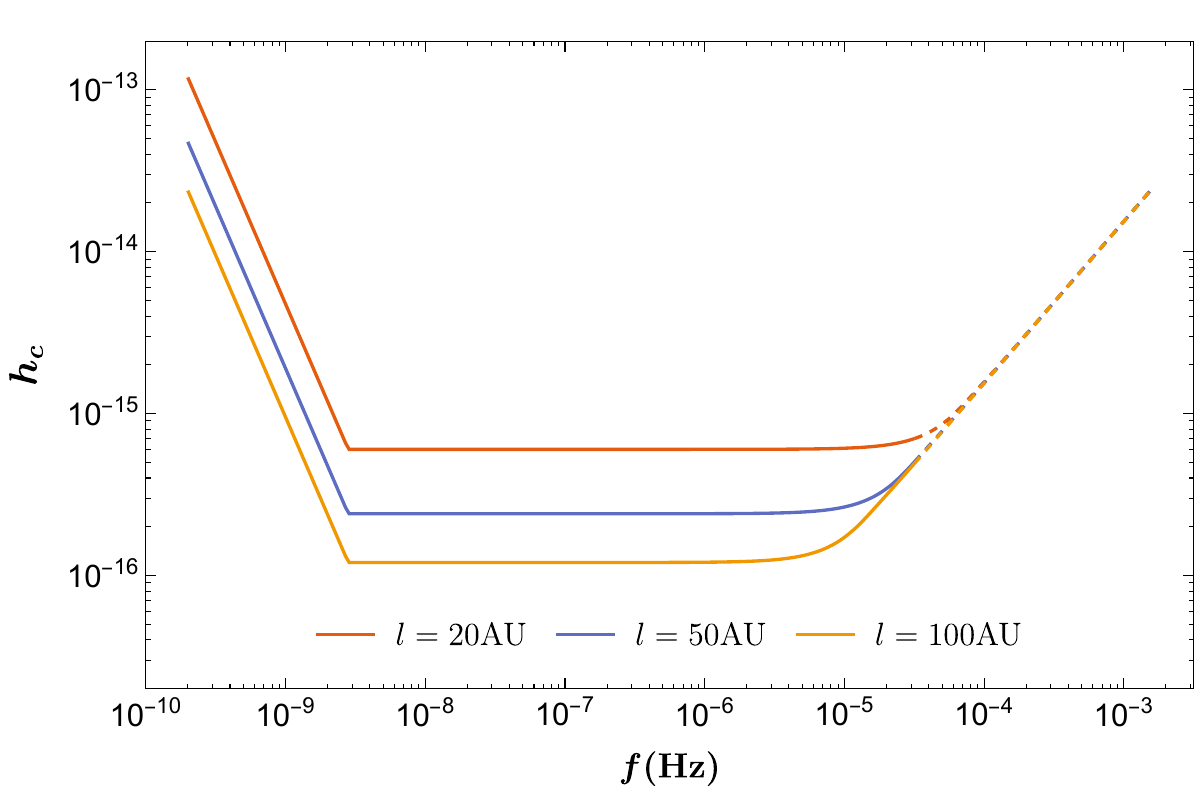}
    \caption{Estimation of the sensitivity on GW strain using auto-correlation. This plot is similar to the result in Fig.~\ref{fig:fullnomodel} but here we have converted our result of $\Omega_{\rm gw}$ into the characteristic strain $h_c$.
    We are assuming 1000 events detected per year and 10 years of observation. We use solid lines in the actual frequency domain of measurement, and dashed lines for the extension into higher frequency, which is only possible with observations of FRB repeaters with a higher detection rate. }
    \label{app:fig:auto_summed}
\end{figure}
The sensitivity curve is shown in Fig. \ref{app:fig:auto_summed}. Notice that this equation only applies within the frequency range $2\pi/T<\omega<2\pi N/T$,  which is roughly the frequency band where the Fourier transform is sensible. For lower frequencies ($\omega<2\pi/T$), one could look for the GW signal as the time derivatives on the arrival time difference, as discussed in Appendix~\ref{app:derivatives}. The parameters are set as: $\delta t_m=0.1$ns, $N=10^4$, $T=10$yr and $N_{\mathrm{FRB}}=1$, where $\delta t_m=0.1$ is the timing precision for each individual FRB event, $N$ is the number of repeating events from one FRB source, and $N_{\mathrm{FRB}}$ is the number of FRB sources.

\subsection*{Angular Correlation Between FRBs}
Another method is to fix the position of one FRB, and analyze the dependency of the correlation function over the other. If our detector can find $10^3$ FRB sources in the sky and these FRBs distribute evenly in the sky, there would always be some FRB located at $\theta_1\approx 0$.  On average, one FRB occupies $4\pi\times 10^{-3}$ solid angle, corresponding to an average angular separation of $6\times 10^{-2}$ between the sources, so we may expect that there exists an FRB source with $\theta_1\sim 10^{-2}$. Thus, it is justified to set $\theta_1=0$ for a crude estimation. In this case, the system has a rotational symmetry around the $x$ axis, and the correlation function should be independent of the $\phi_2$ coordinate. Therefore,  the angular part of the correlation function will depend on $\theta_2$ only with such selection.

Now the correlation function is
\begin{equation}
    C_{\mathrm{angular}}(t,\theta_2)=\int_{-\infty}^\infty\d f e^{i2\pi ft} \frac{S_h(f)}{32\pi^2 f^2}\alpha_{\mathrm{angular}}(\Delta,\theta_2)
\end{equation}
where
\begin{align}
     & \alpha_{\mathrm{angular}}(\Delta,\theta_2)=\int\d\Omega\frac{\sin\left(\frac{\Delta}{2}(1+n_x)\right)\sin\left(\frac{\Delta}{2}(\cos\theta_2+n_x)\right)}{\pi(1+\hat{n}\cdot \hat{s_1})(1+\hat{n}\cdot \hat{s_2})}\nonumber                          \\
     & \times \left\{\left[(\hat{s}_1\cdot\hat{l})^2-(\hat{s}_1\cdot\hat{m})^2\right]\left[(\hat{s}_2\cdot\hat{l})^2-(\hat{s}_2\cdot\hat{m})^2\right]+4(\hat{s}_1\cdot\hat{l})(\hat{s}_1\cdot\hat{m})(\hat{s}_2\cdot\hat{l})(\hat{s}_2\cdot\hat{m})\right\}
\end{align}

\begin{figure}
    \centering
    \subfloat[$\Delta=1$]{\includegraphics[width=0.48\linewidth]{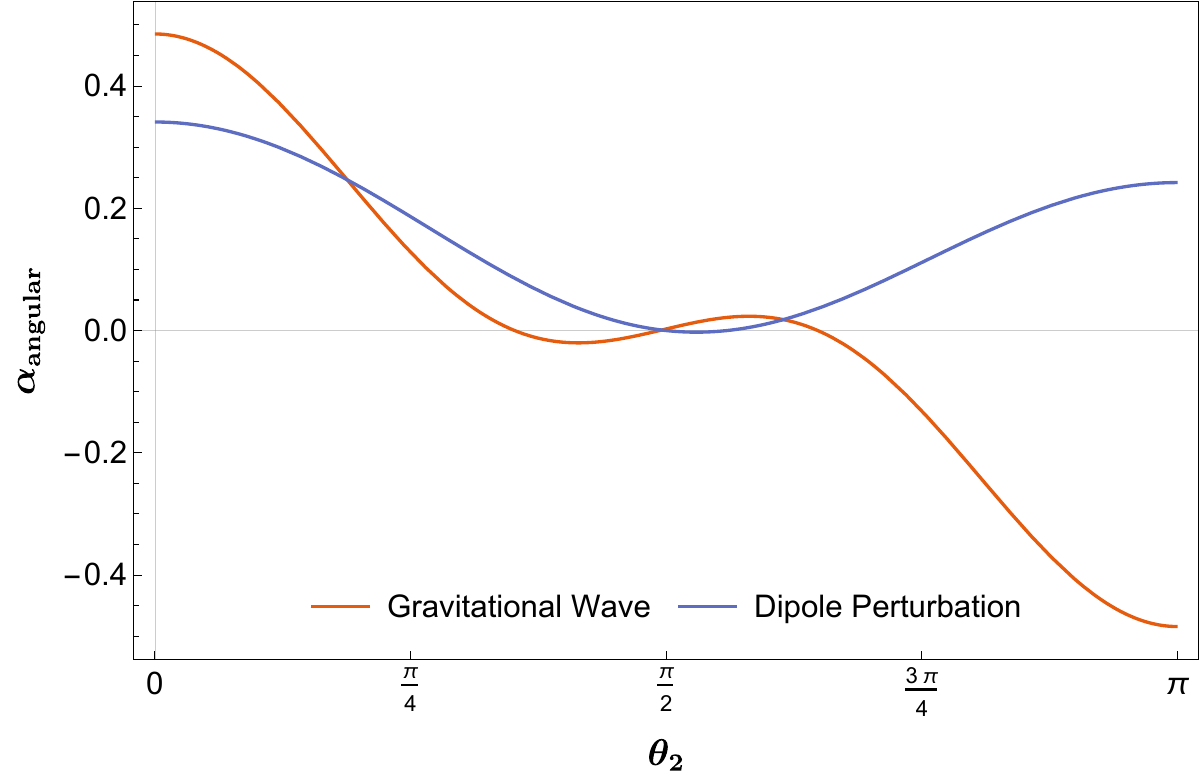}\label{app:fig:ang_d=1}}
    \subfloat[$\Delta=5$]{\includegraphics[width=0.48\linewidth]{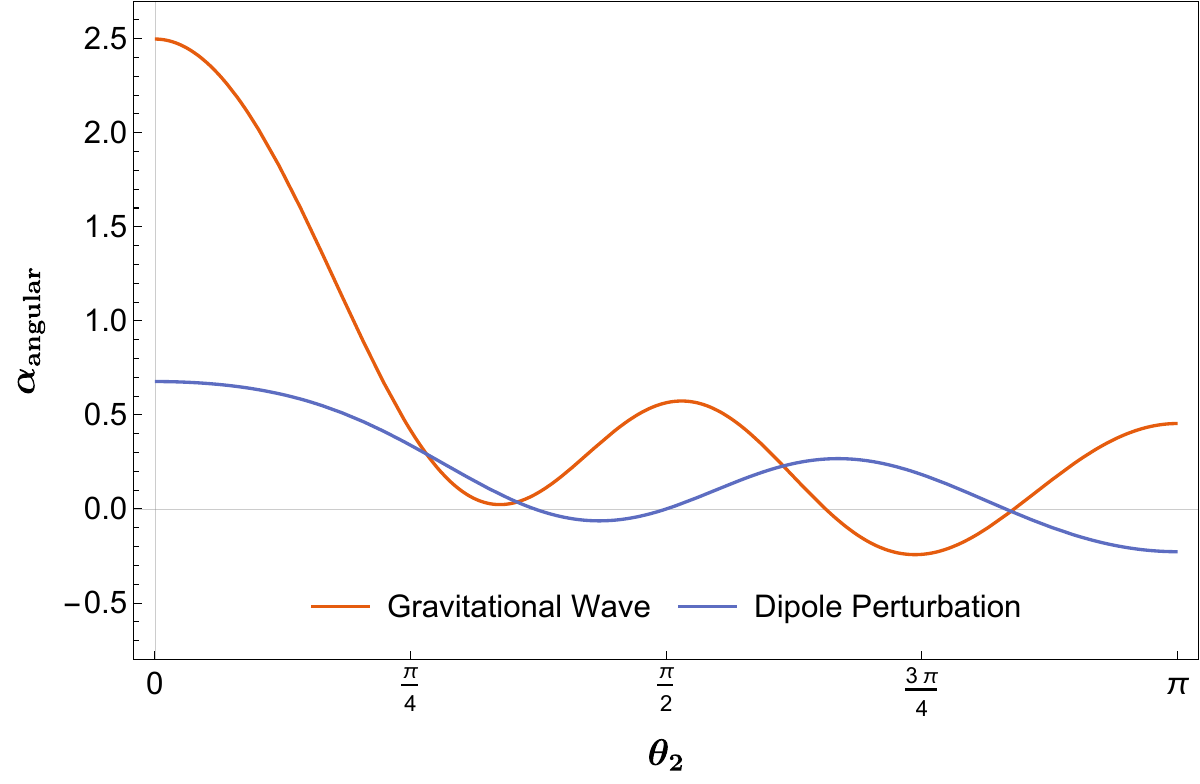}\label{app:fig:ang_d=5}}

    \subfloat[$\Delta=8$]{\includegraphics[width=0.48\linewidth]{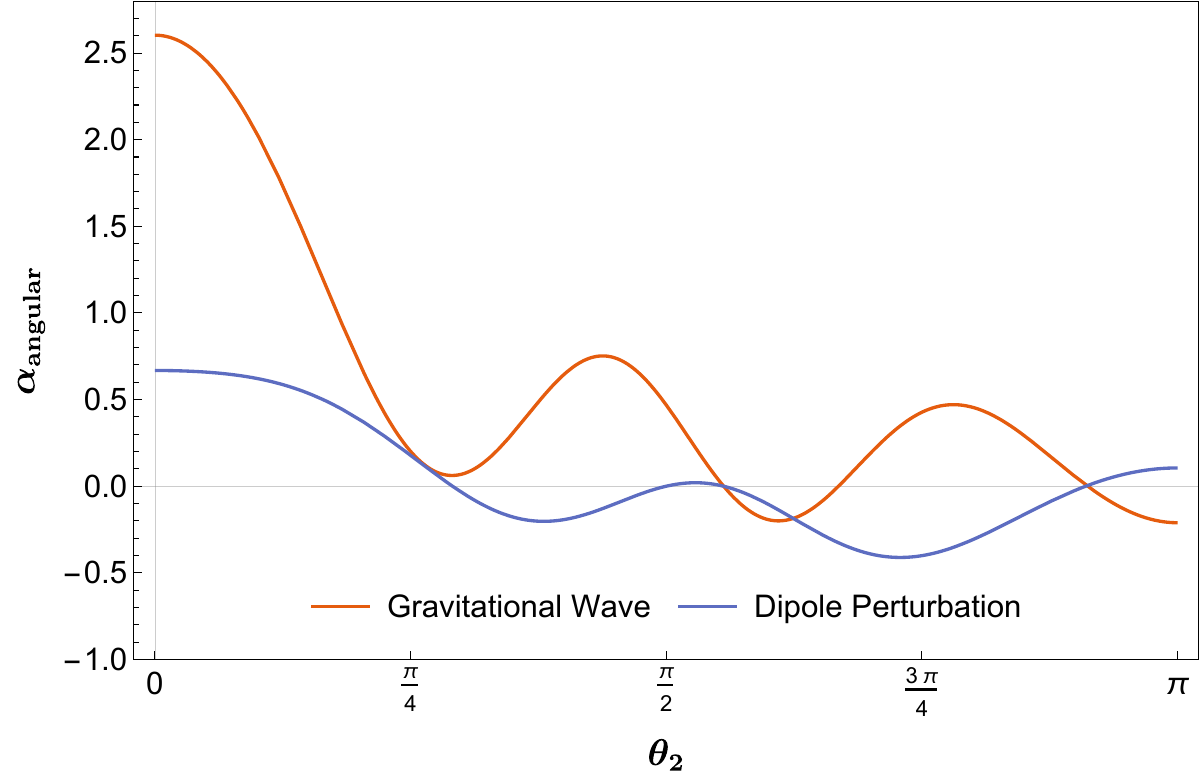}\label{app:fig:ang_d=8}}
    \subfloat[$\Delta=15$]{\includegraphics[width=0.48\linewidth]{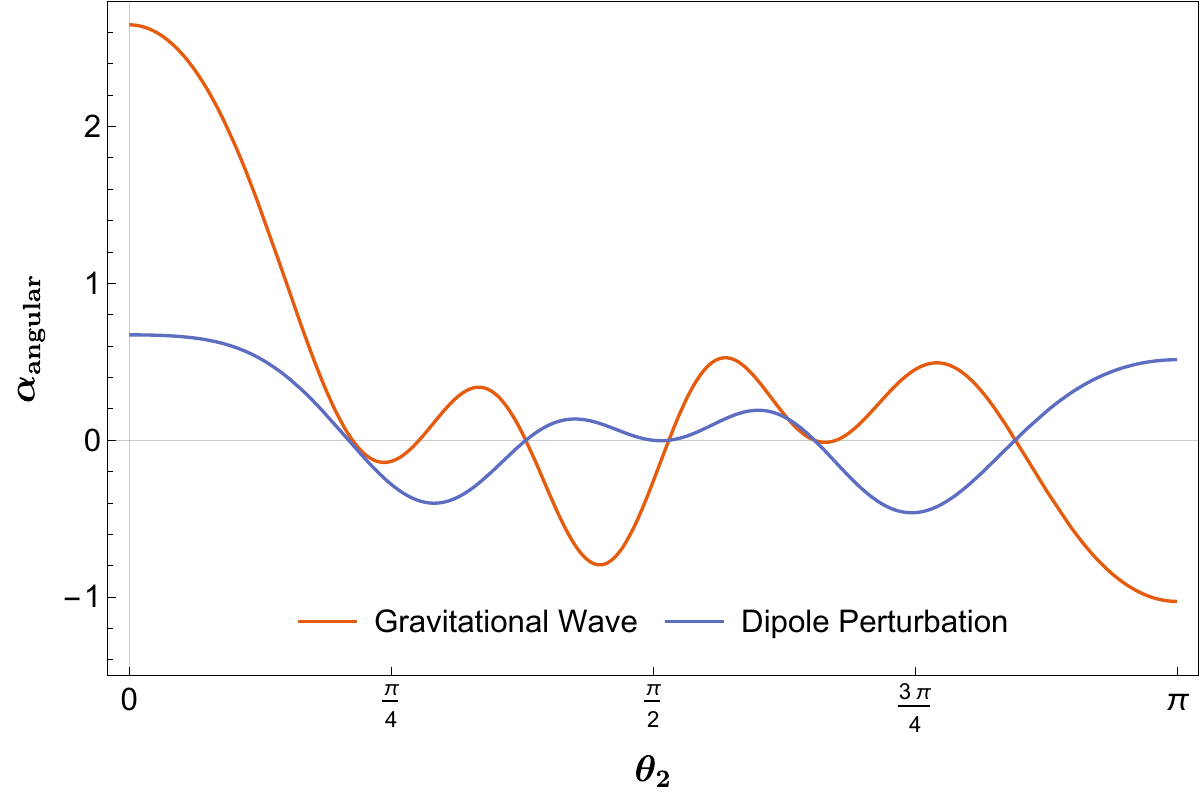}\label{app:fig:ang_d=15}}
    \caption{In this figure, we present $\alpha_{\mathrm{angular}}$ at different frequencies ($\Delta=f\,l=1,5,8,15$), which represents the angular dependence of the frequency space correlation function. The angular factor of gravitational wave signal is shown by the orange line, while the angular factor of dipole perturbation signal is shown by the blue line. Here the direction of one FRB source is fixed to be along the direction of the two dishes. }\label{app:fig:ang}
\end{figure}

The analytic form of $\alpha_{\mathrm{angular}}$ is extremely complicated, involving many special functions. The numerical results are presented in \ref{app:fig:ang} where the angular factor of gravitational wave signal is shown together with the angular factor of dipole perturbation signal, in order to make comparison. We can see that the angular dependence of the correlation function can distinguish between gravitational wave signal and dipole perturbation signal.
It is useful to expand the result in the small $\Delta$ limit. In the setup of this paper, $\Delta \lesssim 1$, and the expansion to $\mathcal{O}(\Delta^4)$ makes a very good approximation.
\begin{equation}
    \alpha_{\mathrm{angular}}(\Delta,\theta_2)\approx\frac{\Delta^2}{30}(11\cos\theta_2+5\cos3\theta_2)+\frac{\Delta^4}{40320}(-14-1362\cos\theta_2-56\cos2\theta_2-637\cos3\theta_2+70\cos4\theta_2-49\cos5\theta_2)
\end{equation}

\section{Variance of the Arrival Time Difference}\label{app:derivatives}

At low frequencies, the observation time cannot encompass one period of gravitational wave. In this case, we measure the time variation of arrival time difference to place limits on the GW signal.
In the low-frequency limit, the arrival time difference $\Delta t(t)$ can be approximated by a Taylor series,
\begin{equation}
    \Delta t(t)=a_0+a_1 t+a_2 t^2+\cdots
\end{equation}
The first term is time-independent, and will be subtracted from the data; the second term is linear in $t$, which is degenerate with the contributions from large-scale structures \cite{Xiao:2024qay}. Therefore, the coefficient $a_2$ can be used to measure or constrain the gravitational wave background. 

In the low frequency regime, equation \eqref{app:equ:dt} becomes
\begin{equation}
    \Delta t(t)=\int\d\Phi \left[1+(2\pi ft)+\frac{1}{2}(2\pi ft)^2+\cdots\right]l\left(\cos\theta_0+n_x\right)\frac{\hat{s}_a\hat{s}_b\tilde{h}_{ab}}{2(1+\hat{n}\cdot\hat{s})}
\end{equation}
so we can extract $a_2$ from this equation
\begin{equation}
    a_2=\int\d\Phi \pi^2f^2l(\cos\theta_0+n_x)\frac{\hat{s}_a\hat{s}_b\tilde{h}_{ab}}{1+\hat{n}\cdot\hat{s}}
\end{equation}
The variance of $a_2$ is,
\begin{equation}
    \left<(a_2)^2\right>=\int_0^\infty\d f\int\d^2\hat{n}\frac{3\pi fl^2}{16}H_0^2\Omega_{gw}(f)(\cos\theta_0+n_x)^2(1-\hat{n}\cdot\hat{s})^2
\end{equation}
Integrate out the angular part,
\begin{equation}
    \left<(a_2)^2\right>=\int_0^\infty\d f\frac{\pi^2}{20}fl^2H_0^2\Omega_{gw}(f)(7+\cos2\theta_0)
\end{equation}
Similar as before, we can add up the contribution of all FRB sources, the effect is to average $\left<(a_2)^2\right>$ over the source position, and divide by $N_{\mathrm{FRB}}$ the final sensitivity. Averaging over source position, we have
\begin{equation}
    \left<(a_2)^2\right>=\int_0^\infty\d f\frac{\pi^2}{3}fl^2H_0^2\Omega_{gw}(f)
\end{equation}
The sensitivity in measuring $a_2$ can be estimated by
\begin{equation}
    (a_2)^2\sim\frac{480\delta t_m^2}{N_{\mathrm{FRB}}NT^4},
\end{equation}
which is the error of fitting a quadratic polynomial. So overall, the sensitivity to $\Omega_{gw}(f)$ per logarithm $f$ is
\begin{equation}
    \Omega_{gw}(f)=\frac{1440\delta t_m^2}{\pi^2N_{\mathrm{FRB}}Nf^2l^2H_0^2T^4}
\end{equation}
This sensitivity is a few orders of magnitude better than required to probe the stochastic GW signal detected by NANOGRAV even at lower frequencies. Therefore, this proposal could not only fill the $\mu$Hz but also improve the measurement at nHz frequencies.

\bibliographystyle{apsrev4-2}
\bibliography{ref}
\end{document}